\documentclass{easychair}

\usepackage{makeidx}  
\usepackage{color}
\usepackage{amsmath}
\usepackage{amssymb}
\usepackage{multirow}
\usepackage{dsfont}
\usepackage{graphicx}

\def\Angstrom{$\mathring{A}$}

\begin{document}


\title{Lattice model refinement of protein structures}
\titlerunning{Lattice model refinement of protein structures}

\author{Martin Mann\\
	Bioinformatics, University of Freiburg,
	Germany,\\
	\url{mmann@informatik.uni-freiburg.de}\\
	\and\\
	Alessandro Dal Pal\`{u}\\
	Dip. di Matematica, Universit\`{a} di Parma, Italy,\\
	\url{alessandro.dalpalu@unipr.it}
	}
\authorrunning{Mann \& Dal Pal\`{u}}


\maketitle              


\begin{abstract}
To find the best lattice model representation of a given full atom protein
structure is a hard computational problem. Several greedy methods have been
suggested where results are usually biased and leave room for improvement.

In this paper we formulate and implement a Constraint Programming method to
refine such lattice structure models. We show that the approach is able to
provide better quality solutions. The prototype is implemented in COLA and is
based on limited discrepancy search. Finally, some promising extensions based on
local search are discussed.
\end{abstract}

\section{Introduction}

Extensive structural protein studies are computationally not feasible using full
atom protein representations. The challenge is to reduce complexity while
maintaining detail \cite{Dill_2008,Istrail_2009}. Lattice protein models are
often used to achieve this but in general only the protein backbone or the amino
acid center of mass is represented
\cite{Backofen_Will_Constraints2006,Mann_LatPack_HFSP_08,Mann:Will:Backofen:CPSP-tools:BMCB:2008,Miao_hydroCollapse_JMB_04,Bornberg:97a}.
A huge variety of lattices and energy functions have previously been developed
\cite{Dill_1985,Godzik_backboneFit_93,Reva_1996}, while the lattices 2D-square,
3D-cubic and 3D face centered cubic (FCC) are most prominent.

In order to evaluate the applicability of different lattices and to enable the
transformation of real protein structures into lattice models, a representative
lattice protein structure has to be calculated. In detail, given a full atom
protein structure one has to find the best structure representation within the
lattice model that minimizes the applied distance measure. Ma\v{n}uch and Gaur
have shown the NP-completeness of this problem for backbone-only models in the
3D-cubic lattice when minimizing coordinate root mean square deviation (cRMSD)
and named it the \emph{protein chain lattice fitting (PCLF)
problem}~\cite{Manuch_2008}.

The PCLF problem has been widely studied for backbone-only models.
Suggested approaches utilize quite different methods, ranging from full
enumeration~\cite{Covell_1990}, greedy chain growth
strategies~\cite{Mann_latfit_2010,Miao_hydroCollapse_JMB_04,Park:Levitt:JMB1995},
dynamic programming~\cite{Hinds_1992}, simulated
annealing~\cite{Ponty_backboneFits_NAR_08}, or the optimization of specialized
force fields~\cite{Koehl_1998,Reva_1998}. The most important aspects in
producing lattice protein models with a low root mean squared deviation (RMSD)
are the lattice co-ordination number and the neighborhood vector angles
\cite{Park:Levitt:JMB1995,Pierri_Proteins_08}. Lattices with intermediate
co-ordination numbers, such as the face-centered cubic (FCC) lattice, can produce
high resolution backbone models \cite{Park:Levitt:JMB1995} and have been used in
many protein structure studies (e.g.
\cite{Istrail_2009,Jacob_2007,Ullah:CPSP_LS:09}).

Most of the PCFL methods introduced are heuristics to derive good solutions in
reasonable time. Greedy methods as chain growth
algorithms~\cite{Mann_latfit_2010,Miao_hydroCollapse_JMB_04,Park:Levitt:JMB1995}
enable low runtimes but the fitting quality depends on the chain growth direction and
parameterization. Thus, resulting lattice models are biased by the method
applied and have potential for refinement.

This paper has the goal to provide some evidence that greedy methods
can be effectively improved by subsequent refinement steps that increase the
fitting quality. We present a formalization and a simple working prototype.
Moreover we briefly discuss some potential methodologies that we expect could be
effectively employed.

\section{Definitions and Preliminaries}
\label{sec:pre}

\newcommand{\dN}{\text{neigh}}
\newcommand{\CA}{$C_{\alpha}$}
\newcommand{\DIST}{\operatorname{dist}}

In order to define the Constraint Programming approach we first introduce some
preliminary formalisms.

Given a protein in full atom representation of length~$n$ (e.g. in Protein
Data Base (PDB) format~\cite{PDB_2000}), we denote the sequence of
3D-coordinates of its \CA-atoms (its \emph{backbone trace}) by $P=(P_1,\ldots,P_n)$.

A regular \emph{lattice~$L$} is defined by a set of neighboring
vectors~$\vec{v}\in N_L$ of equal length $(\forall_{\vec{v_i},\vec{v_j}\in N_L}
: |\vec{v_i}|=|\vec{v_j}|)$, each with a reverse $(\forall_{\vec{v}\in N_L} :
-\vec{v}\in N_L$, such that $L = \{\vec{x} \;|\; \vec{x} = \sum_{\vec{v}_i\in
N_L} d_i \cdot \vec{v}_i \wedge d_i \in \mathds{Z}^{+}_{0} \}$. $|N_L|$~gives
the coordinate number of the lattice~$L$, e.g. 6~for 3D-cubic or 12~for the FCC
lattice. All neighboring vectors $\vec{v} \in N_L$ of the used lattice $L$ are
scaled to a length of~3.8\Angstrom{}, which is the mean distance between
consecutive \CA-atoms in real protein structures.

A backbone-only \emph{lattice protein structure~$M$} of length~$n$ is defined by
a sequence of lattice nodes $M = (M_1,\ldots,M_n) \in L^n$ representing the
backbone (\CA) monomers of each amino acid. A valid structure ensures backbone
connectivity $(\forall_{i<n} : M_i-M_{i+1} \in N_L)$ as well as selfavoidance
$(\forall_{i\neq j} : M_i \neq M_j)$, i.e. it represents a selfavoiding walk
(SAW) in the underlying lattice.

The \emph{PCFL problem} is to find a lattice protein model~$M$ of a given
protein's backbone~$P$, such that a distance measure between~$M$ and~$P$
($\DIST(M,P)$) is minimized~\cite{Manuch_2008}.

In this contribution, we tackle the \emph{PCFL refinement problem}. Here, a
protein backbone~$P$ as well as a first lattice model~$M$ is given, e.g. derived
by a greedy chain growth
procedure~\cite{Mann_latfit_2010,Miao_hydroCollapse_JMB_04,Park:Levitt:JMB1995}.
The problem is to find a lattice model~$M'$, such that $\DIST(M',P) <
\DIST(M,P)$, via a relaxation/refinement of the original model~$M$.

In the following, we utilize distance RMSD (dRMSD, Eq.~\ref{eq:dRMSD}) as the
distance measure~$\DIST(M,P)$. dRMSD is independent of the relative orientation
of~$M$ and~$P$ since it captures the model's deviation from the pairwise
distances of \CA{}-atoms in the original protein. Minimizing this measure
optimizes the lattice model obtained.
\begin{eqnarray}
	\text{dRMSD}(M,P) &=& \sqrt{ \frac{ \sum_{i<j} \; (|M_j-M_i| - |P_j-P_i|)^2 }{
	n(n-1)/2 } } \label{eq:dRMSD}
\end{eqnarray}

\section{Refinement of Lattice Models: a Constraint Model in COLA}

In this section we formalize a Constraint Optimization Problem (COP) to solve
the PCFL refinement problem (see Sec.~\ref{sec:pre}), i.e. to refine a lattice
model~$M$ of a protein~$P$. The input is the original protein~$P$ and its lattice model~$M$
to be refined. The output is a lattice model~$M'$ derived from~$M$ via some
relaxation that optimizes our distance measure dRMSD$(M',P)$
(Eq.~\ref{eq:dRMSD}).

We first formalize the problem and show how to implement it in COLA, a
COnstraint solver for
LAttices~\cite{Dal_Palu:Dovier:Fogolari:Const_Logic_Progr:2004}. This is
followed by an altered formulation that utilizes limited discrepancy
search~\cite{Harvey95limiteddiscrepancy}.


\newpage

\subsection{The Constraint Optimization Problem}

The COP can be formalized as follows:

\begin{center}
\begin{tabular}{p{0.25\textwidth} p{0.7\textwidth}}
	$X_1 \ldots X_n$ & variables representing $M' = (M'_1,\ldots,M'_n)$ \\
\\
	$D(X_i)$ & variable domains $= \{ v \;|\; v \in L \wedge | v -
	M_i | \leq f_{\text{scale}} \cdot d_{\max} \}$, \\
	& i.e. an $M_i$ surrounding sphere with
	radius $f_{\text{scale}}\cdot d_{\max}$ \\
\\
	$SAW(X_1 \ldots X_n)$ & self-avoiding walk constraint, e.g. split into
	a chain of binary \texttt{contiguous} and a global \texttt{alldifferent}
	constraint\\
\\
	$O$ & objective function variable, implements dRMSD\\
	 & $ = \sum_{i<j} (|X_j - X_i| - |P_j - P_i|)^2$
	 to be minimized\\
\end{tabular}
\end{center}

Note that $d_{\max}$ refers to the number of lattice units used and thus it is
scaled to the correct distance of $f_{\text{scale}}=3.8$\AA{}. Thus, the
domains for $d_{\max}=0$ only contain the original lattice point~$M_i$ (domain size~1),
while $d_{\max}=1$ results in~$M_i$ as well as all neighbored lattice points
(domain size~$1+12=13$ in FCC). The domain size guided by $d_{\max}$ defines the
allowed relaxation of the original lattice model~$M$ to be refined. For more
details about global constraints for protein structures on lattices, the reader
can refer to~\cite{Backofen_Will_Constraints2006,DalPalu_2010}.

The COLA implementation takes advantage of the availability of 3D lattice point
domains and distance constraints. The implementation changes the original
framework only in the input data handling and objective function definition. A
working copy of COLA and the COP implemented for this paper are available
at \texttt{http://www2.unipr.it/$\sim$dalpalu/COLA/}

\subsection{Limited Discrepancy Search}

A simple enumeration with $d_{\max}=1$ and a protein of length~50, already shows
that the search space of the COP from the previous section is not manageable. In
this example, each point can be placed in 13~different positions in the FCC
lattice, and even if the contiguous constraint among the amino acids is
enforced, the number of different paths is still beyond the current
computational limits.

We tried a simple branch and bound search an $X_1,\ldots,X_n$, where the dRMSD
bound is estimated by considering the possible placement of non labeled
variables and the best dRMSD contribution provided by each amino acid. In
detail, each amino acid $s$ not yet labeled is compared to each other amino acid
($s'$). Each pair provides a range of different contributions to dRMSD measure,
depending on the placement of $s$ and the placement of the other amino acids
(when not yet labeled). A closed formula computation (rather than a full
enumeration of all combinations), based on bounding box of domain positions, is
activated, in order to estimate the minimal contribution. Clearly, this
estimation is not particularly suited, since we relax the estimation on
$\mathds{R}^3$, where the null (best) contribution can be easily found as soon
as the bounds on $|X_s-X_{s'}|$ include the value $|P_s-P_{s'}|$. Unfortunately,
the discrete version requires a more expensive evaluation that boils down to
full pair checks. Therefore, the current bound is very loose and the pruning
effects are modest.

A general impression is that the dRMSD measure presents a pathological
distribution of local minima, depending on the placement of amino acids on the
lattice. In general, due to the discrete nature of the lattice, the modification
of a single amino acid's position can  drastically vary its contributions to
the measure.

These considerations suggested us to focus on the identification of solutions
that improve the dRMSD w.r.t. $M$ rather than searching for the optimal one.
In terms of approximated search we tried to capture the main characteristics of
the COP and design efficient and effective heuristics.

A simple idea we tested is the \emph{limited discrepancy}
search~\cite{Harvey95limiteddiscrepancy}. This search compares the amino acid
placements in the lattice models~$M$ and~$M'$. Every time a corresponding amino
acid is placed differently in the two conformations, we say that there is a
\emph{discrepancy}. We set a global constraint that limits the number of
deviations to at most~$K$. This allows to generate conformations that are rather
similar to~$M$, especially if $d_{\max}$ is greater than~1. The rational behind
this heuristics is that we expect that potential conformations $M'$ improve the
dRMSD only when contained in a close neighborhood of the $M$ structure.

The count of the number of discrepancies $K$ is implemented directly in COLA at
each labeling step.

\subsection{Results}

We summarize here the preliminary results coming from the COLA implementation of
a $K$ discrepancy search in 3D FCC lattice.

The initial lattice models to be refined were generated using the LatFit tool
from the LatPack package~\cite{Mann_LatPack_HFSP_08,Mann_latfit_2010}. LatFit
implements an efficient greedy dRMSD optimizing chain growth method and was
parameterized to consider the best 100~structures from each elongation for
further growth\footnote{For details on the LatFit method see
\cite{Mann_latfit_2010} and the freely available web interface at
{\small\texttt{\url{http://cpsp.informatik.uni-freiburg.de}}}}.

\begin{table}[tb]
\centering
\begin{tabular*}{0.5\textwidth}{@{\extracolsep{\fill}}l|c|c|c|}
Protein ID	& 8RXN	& 1CKA	& 2FCW
\\
\hline
length		& 52	& 57	& 106
\\
\end{tabular*}
\vspace{1em}
\caption{Used proteins from the Protein Data Base (PDB)~\cite{PDB_2000}.}
\label{tab:ID}
\vspace{-1em}
\end{table}

We test three proteins (Table~\ref{tab:ID}) and for each of them we input the
conformation~$M$ obtained from the greedy algorithm (LatFit).
Table~\ref{tab:results} reports the best dRMSD of our new model~$M'$ found
depending on $d_{\max}$ and the number $K$ of amino acids placed differently
from the input conformation. Furthermore, time consumption for each
parameterization is given.

Note that if either $K=0$ or $d_{\max}=0$ only the input structure resulting
from the greedy LatFit run can be enumerated.

\begin{figure}[tb]
	\centering
	 \includegraphics[width=0.5\textwidth]{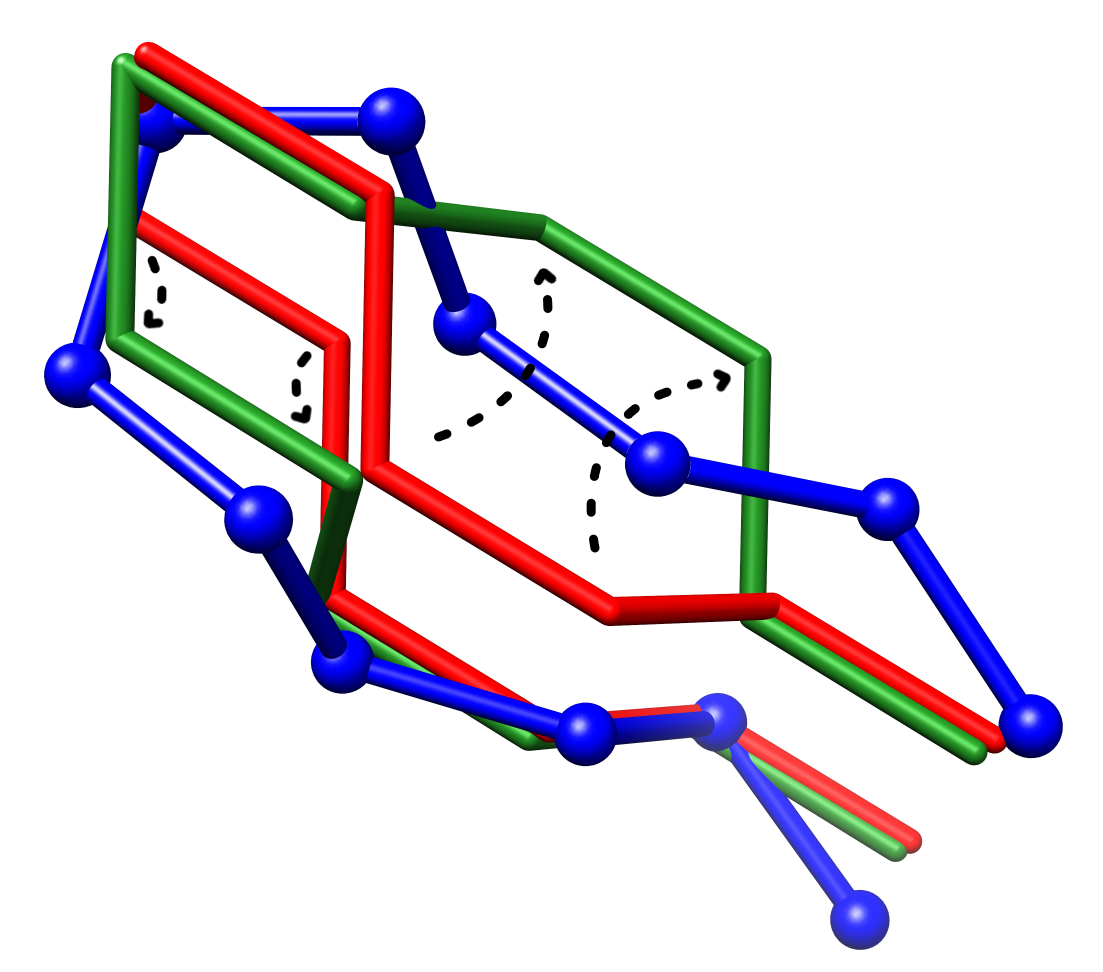}
	\caption{The initial lattice model~$M$ (red) of the
	protein chain~$P$ (blue, balls) and the final/refined lattice model~$M'$
	(green) resulting from $d_{\max}=2$ and $K=4$ for protein \texttt{8RNX}. Note, only the altered
	loop regions (residue 2-14) are shown, but the whole structure models~$M$ and
	$M'$ were superpositioned to~$P$ independently.}
	\label{fig:8RNX-refinement-LDS-3-4}
\end{figure}

\begin{table}[tb]
\centering
\small
\begin{minipage}{0.48\textwidth}
\begin{tabular*}{0.95\textwidth}{@{\extracolsep{\fill}}rr|rrrr|}
\multicolumn{2}{c}{}& \multicolumn{4}{c}{dRMSD}\\
\hline
\multicolumn{2}{|c|}{}& \multicolumn{4}{c|}{$K$}\\
\multicolumn{2}{|c|}{8RXN}& 1 & 2 & 3 & 4 \\
\hline
\multirow{4}{*}{$d_{\max}$}
&$0$ & 1.2469 & 1.2469 & 1.2469 & 1.2469\\
&$1$ & 1.2319 & 1.2172 & 1.1639 & 1.1189\\
&$2$ & 1.2319 & 1.1674 & 1.1596 & 1.0884\\
&$3$ & 1.2319 & 1.1674 & 1.1596 & 1.0884\\
\hline
\multicolumn{2}{|c|}{}& \multicolumn{4}{c|}{$K$}\\
\multicolumn{2}{|c|}{1CKA}& 1 & 2 & 3 & 4 \\
\hline
\multirow{4}{*}{$d_{\max}$}
&$0$ & 1.2370 & 1.2370 & 1.2370 & 1.2370\\
&$1$ & 1.2226 & 1.2226 & 1.2226 & 1.2226\\
&$2$ & 1.2026 & 1.1887 & 1.1887 & 1.1887\\
&$3$ & 1.2026 & 1.1887 & 1.1887 & 1.1887\\
\hline
\multicolumn{2}{|c|}{}& \multicolumn{4}{c|}{$K$}\\
\multicolumn{2}{|c|}{2FCW}& 1 & 2 & 3 & 4 \\
\hline
\multirow{4}{*}{$d_{\max}$}
&$0$ & 1.1353 & 1.1353 & 1.1353 & 1.1353 \\
&$1$ & 1.1353 & 1.1324 & 1.1317 & 1.1309 \\
&$2$ & 1.1321 & 1.1300 & 1.1254 & 1.1200 \\
&$3$ & 1.1321 & 1.1300 & 1.1254 & 1.1200 \\
\cline{3-6}
\end{tabular*}
\end{minipage}
\begin{minipage}{0.48\textwidth}
\begin{tabular*}{0.95\textwidth}{@{\extracolsep{\fill}}rr|rrrr|}
\multicolumn{2}{c}{}& \multicolumn{4}{c}{time in seconds}\\
\hline
\multicolumn{2}{|c|}{}& \multicolumn{4}{c|}{$K$}\\
\multicolumn{2}{|c|}{8RXN}& 1 & 2 & 3 & 4 \\
\hline
\multirow{4}{*}{$d_{\max}$}
&$0$ & 0.048 & 0.081 & 0.040 & 0.039 \\
&$1$ & 0.112 & 0.790 & 2.365 & 20.70 \\
&$2$ & 0.068 & 0.983 & 6.500 & 106.6 \\
&$3$ & 0.106 & 0.499 & 7.399 & 124.0 \\
\hline
\multicolumn{2}{|c|}{}& \multicolumn{4}{c|}{$K$}\\
\multicolumn{2}{|c|}{1CKA}& 1 & 2 & 3 & 4 \\
\hline
\multirow{4}{*}{$d_{\max}$}
&$0$ & 0.031 & 0.030 & 0.027 & 0.037 \\
&$1$ & 0.402 & 0.615 & 3.442 & 39.27 \\
&$2$ & 0.225 & 0.456 & 7.595 & 120.6 \\
&$3$ & 0.421 & 0.616 & 8.573 & 140.2 \\
\hline
\multicolumn{2}{|c|}{}& \multicolumn{4}{c|}{$K$}\\
\multicolumn{2}{|c|}{2FCW}& 1 & 2 & 3 & 4 \\
\hline
\multirow{4}{*}{$d_{\max}$}
&$0$ & 0.043 & 0.050 & 0.058 & 0.078 \\
&$1$ & 0.118 & 1.997 & 49.99 & 1128 \\
&$2$ & 0.294 & 7.192 & 341.8 & 14235 \\
&$3$ & 0.332 & 8.129 & 394.5 & 16140 \\
\cline{3-6}
\end{tabular*}
\end{minipage}
\\
\vspace{1em}
\caption{$d_{\max}$ and $K$ influence on discrepancy search measured in
dRMSD and time.}
\label{tab:results}
\vspace{-1em}
\end{table}

These results, yet preliminary, offer an interesting insight about the
distribution of suboptimal solutions. It is interesting to note, e.g., that
better solutions are found by allowing a rather large local neighborhood for a
few amino acids ($d_{\max}$ parameter). On the other side, it seems that few
modifications ($K$) are sufficient to alter the input sequence and obtain a
better conformation.

In Figure~\ref{fig:8RNX-refinement-LDS-3-4} we exemplify the gain of model
precision for the protein \texttt{8RNX}. Only the relaxation of $K=4$ monomers
enables the structural change that leads to a dRMSD drop from 1.2469 down to
1.0884, an improvement of about~13\%. A movement of less monomers would not
enable such a drastic change. This depicts the potential of a local search
scheme that iteratively applies a series of such structural changes.

Investigating the time consumption (Table~\ref{tab:results}) one can see that
the runtime increases drastically with $K$ which governs the search tree size.
The domain sizes implied by $d_{\max}$ do not show such an immense influence.

The behavior encountered is an indicator that a search based on exploring only
the neighborhood should provide efficient and good suboptimal solutions. In the
next section we briefly discuss some promising approaches that we plan to
investigate.

\clearpage
\newpage

\subsection{Future work}

In our opinion, a framework that integrates CP and  Local Search is particularly
suited to generate fast suboptimal solutions, potentially very close to the
optimal one. We identify some possible directions that we believe are excellent
candidates to model and solve approximately the PCLF problem:

\begin{itemize}
  \item {\bf local neighboring search \cite{Cipriano_GELATO,Dotu:08}}: this
  technique allows to integrate Gecode and Local Search frameworks. The
  framework handles constraint specifications and local moves within C++
  programming language;
  \item {\bf $k$-local moves \cite{Ponty_backboneFits_NAR_08}:} the idea here
  is to apply structural changes on $k$ consecutive amino acids and repeat the
  process in a Monte-Carlo and/or simulated annealing style.
  \item {\bf side chain model \cite{Mann:local_move:WCB09}}: our model can be
  extended to include side chains and we could exploit a similar set of local
  moves.
  \item {\bf the framework presented in~\cite{COLAandLS}}: COLA is here
  extended and combined directly to a Local Search approach based on \emph{pull
  moves} \cite{Lesh:RECOMB2003}.
\end{itemize}


\section{Conclusion}

In this paper we presented a Constraint Programming based model for the
refinement of lattice fitting of protein conformations. A simple branching was
shown to be ineffective and a limited discrepancy search was modeled and shown
to be beneficial to the identification of suboptimal solutions. A prototypical
implementation in the framework COLA and some preliminary results have shown the
feasibility of the method. We believe that an extension of the framework to
Local Search is particularly suited for the PCLF problem at hand.

\paragraph{Acknowledgments} This work is partially supported by PRIN08 \emph{Innovative multi-disciplinary approaches
for constraint and preference reasoning} and GNCS-INdAM
\emph{Tecniche innovative per la programmazione con vincoli in applicazioni strategiche}.

\bibliographystyle{plain}
\bibliography{references}

\end{document}